# The Long-Term X-Ray Lightcurve of RX J0527.8–6954

J. Greiner[1], R. Schwarz[2], G. Hasinger[2], M. Orio[3]*

[1] Max-Planck-Institut für Extraterrestrische Physik, 85740 Garching, Germany
[2] Astrophysikalisches Institut Potsdam, 14482 Potsdam, Germany
[3] Department of Physics, University of Wisconsin, Madison, WI 53706, USA

**Abstract.** Supersoft X-ray sources are commonly believed to be stably burning white dwarfs. However, the observations of some supersoft sources show dramatic variability of their X-ray flux on timescales ranging from days to years. Here, we present further observational data of the supersoft X-ray source RX J0527.8–6954 exhibiting a continuous decline over the past 5 yrs. With no clear trend of a concordant temperature decrease this might suggest a evolutionary scenario where the WD leaves the steady burning branch and the combined effect of reduced luminosity and cooling at constant radius produces the observed effect.

## 1 Introduction

The supersoft X-ray source RX J0527.8–6954 was discovered in the *ROSAT* first light observation (Trümper et al. 1991) of the Large Magellanic Cloud (LMC) in June 1990. It has an extremely soft X-ray spectrum, with spectral parameters (blackbody temperature, absorbing column density) very similar to CAL 83 (Greiner et al. 1991), the SSS prototype (Long et al. 1981). Also, it was readily realized that this source must have brightened up by at least a factor of 10 compared to previous *Einstein* observations when RX J0527.8–6954 was in the field of view but not detected. It was noticed already earlier (Orio and Ögelman 1993, Hasinger 1994) that the countrate of RX J0527.8–6954 had decreased substantially since its discovery.

## 2 Observational Results

### 2.1 All-Sky-Survey

As RX J0527.8–6954 is close to the south ecliptic pole, it was scanned during the All-Sky-Survey over a time span of 21 days. The total observation time resulting from 92 individual scans adds up to 1.96 ksec. Due to the scanning mode the source has been observed at all possible off-axis angles with its different widths of the point spread function. For the temporal and spectral analysis we have used an 5′ extraction radius to ensure that no source photons are missed.

---

* On leave from Osservatorio Astronomico di Torino, 10025 Pino Torinese, Italy

No other source down to the $1\sigma$ level is within this area. Each photon event has been corrected for its corresponding effective area. The background was determined from a circle 13′ off along southern ecliptic latitude with respect to RX J0527.8–6954. The mean countrate was determined to $(0.14\pm0.06)$ cts/sec. Due to systematic errors of about 20% no definite conclusion can yet be drawn on possible short-term variations of the X-ray flux.

For the spectral fitting the X-ray photons in the amplitude channels 11–240 (though there are almost no photons above channel 50) were binned with a constant signal/noise ratio of $5\sigma$. The fit of a blackbody model results in an effective temperature of $kT_{bb} = 40$ eV (with the absorbing column fixed at its galactic value), very similar to the results obtained from fitting the PSPC data of the *ROSAT* first light observation (Greiner et al. 1991).

### 2.2 PSPC Pointings

Several pointings with the *ROSAT* PSPC have been performed on RX J0527.8–6954, starting with the first light observation and continuing with several dedicated pointings. In addition, RX J0527.8–6954 was also in the field of view of a number of other target pointings, mainly within those on the bright supernova remnant N132D which has been used for calibration purposes. We restricted the analysis to pointings with effective exposure times larger than 300 sec and containing RX J0527.8–6954 at less than 50′ off-axis angle (Table 1).

Depending on the off-axis angle of RX J0527.8–6954 within the detector, photons have been extracted within 6–10′ radius. This extreme size of the extraction circle was chosen because the very soft photons (below channel 20) have a much larger spread in their measured detector coordinates. As usual, the background was determined from a ring well outside the source without having contaminating sources in there. Before subtraction, the background photons were normalized to the same area as the source extraction circle. Since RX J0527.8–6954 is affected by the window support structure in many pointings, we have developed a dedicated procedure to correct for the shadowing/wobbling effect.

As a first step, we took the standard PSPC instrument map together with the effective area table to produce an instantaneous correction map. In the second step these correction maps are added up according to the wobble motion and roll-angle using the attitude table. Both these steps were performed separately in the 11–41 and 42–52 energy channel bands. This energy selection is important because the obscuration of sources (scattering) is energy dependent. We then determine the good time intervals (exposure time) and after multiplication get two exposure maps (in the separate energy bands) containing the effects of vignetting and wobbling. As the next step we determine the relative number of counts of RX J0527.8–6954 in the 11–41 and 42–52 bands and use these as weights for summing the two exposure maps. Finally, the resulting exposure map is used to compute the effective exposure time (given in column 5 of Tab. 1) of RX J0527.8–6954 by averaging over the same location and area as source photons have been extracted.

Table 1. Summary of *ROSAT* observations covering RX J0527.8–6954.

| Observation No. | Date | PSPC or HRI | $T_{Nom}$ (sec) | $T_{Eff}$ (sec) | No. of counts | Off-axis angle | Distance to next rib |
|---|---|---|---|---|---|---|---|
| 110173 | June 18, 1990 | P | 2042 | 1418 | 296 | $31'$ | $5'$ |
| 110176 | June 19, 1990 | P | 2168 | 1548 | 292 | $29'$ | $5'$ |
| 110074 | June 20, 1990 | P | 754 | 471 | 72 | $28'$ | $9'$ |
| 110181 | June 21, 1990 | P | 1882 | 1176 | 207 | $46'$ | $15'$ |
| 110090 | June 24, 1990 | P | 457 | 310 | 69 | $25'$ | $3'$ |
| 110234 | July 6, 1990 | H | 779 | 736 | 19 | $13'$ | – |
| 110241 | July 7, 1990 | H | 474 | 460 | 12 | $12'$ | – |
| Survey | Oct. 10–31, 1990 | P | 1965 | 1357 | 189 | $0$–$55'$ | – |
| 141800 | Dec. 11, 1991 | P | 1060 | 541 | 40 | $21'$ | $0'$ |
| 160084 | May 5, 1991 | P | 1757 | 1040 | 100 | $20'$ | $0'$ |
| 300126 | Mar. 5, 1992 | P | 7802 | 4972 | 401 | $22'$ | $0'$ |
| 500004 | Apr. 5, 1992 | P | 1100 | 805 | 47 | $20'$ | $1'$ |
| 400148 | Apr. 6, 1992 | P | 6263 | 6263 | 658 | $1'$ | $20'$ |
| 300172 | May 7–16, 1992 | P | 6371 | 3636 | 403 | $37$ | $1'$ |
| 400238 | Nov. 26, 1992 | H | 4063 | 3788 | 38 | $10'$ | – |
| 400298 | Dec. 6, 1992 | P | 1058 | 1058 | 61 | $1'$ | $20'$ |
| 300172a | Dec. 16–26, 1992 | P | 2996 | 2163 | 120 | $36'$ | $10'$ |
| 400298a | Mar. 11–16, 1993 | P | 7506 | 7506 | 202 | $1'$ | $20'$ |
| 500141 | Apr. 11, 1993 | P | 5259 | 3709 | 110 | $19'$ | $1'$ |
| 141937 | Apr. 16, 1993 | P | 1979 | 1394 | 55 | $21'$ | $0'$ |
| 141506 | June 16, 1993 | P | 706 | 446 | 15 | $21'$ | $0'$ |
| 300172b | June 14–27, 1993 | P | 3882 | 2630 | 157 | $36'$ | $8'$ |
| 141507 | Aug. 24, 1993 | P | 1334 | 940 | 25 | $22'$ | $0'$ |
| 201689 | Aug. 29/30, 1994 | H | 8463 | 8463 | 16 | $0'$ | – |
| 201996 | Aug. 10–12, 1995 | H | 7162 | 7162 | 8 | $0'$ | – |
| 600782 | Oct. 22-24, 1995 | H | 12872 | 12588 | 6 | $4'$ | – |

## 2.3 HRI Pointings

There are also a number of HRI pointings which cover RX J0527.8–6954, namely two pointings during the verification phase, one pointing performed in November 1992 in the framework of LMC X-ray source identifications (see Cowley et al. 1993) and three dedicated, on-axis pointings in August 1994, August 1995 and October 1995 (see Tab. 1). We restricted the analysis to pointings with effective exposure times longer than 300 sec and at less than $15'$ off-axis angle (excluding two verification phase pointings). Source photons have been extracted within $2$–$3'$ and were background and vignetting corrected.

We derive a best-fit position ($\pm 5''$) of R.A. (2000.0) = $05^{\rm h}27^{\rm m}48\overset{\rm s}{.}9$, Decl. (2000.0) = $-69°54'09''$ which results from the position averaging of the two individual on-axis HRI pointings in August 1994 and August 1995 (which differ by $4''$). The averaging reduces the irreproducible r.m.s. scatter of $\approx 5''$ from the individual pointings due to the fact that the roll angles are different. This new position differs by only 2″.5 from that given in Hasinger (1994) which arose from the averaging of several off-axis (4 PSPC and 1 HRI) pointings. Our new position strengthens the conjecture of Hasinger (1994) that the blue object proposed by Cowley et al. (1993) as a strong counterpart candidate is too distant to be a likely counterpart (see Fig. 1). Though this new position is only 6″ from the Cowley et al. (1993) position, it is in the opposite direction with respect to the above mentioned blue object.

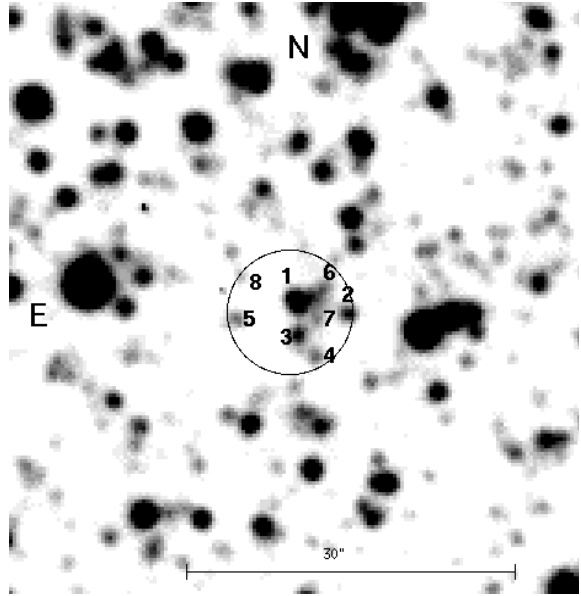

**Fig. 1.** The X-ray error circle of 5″ radius overplotted on a 10 min B image taken on March 25, 1995 with the ESO 2.2 m telescope at La Silla/Chile (Max-Planck-Institute time). All resolved objects inside the error box are numbered. Object 1 has been spectroscopically identified as B8IV (Cowley et al. 1993). A spectrum of object 2 ($m_B$=18$^m$.5) taken on March 28, 1995 (also ESO 2.2 m) suggests spectral type B. Object 6 seems to be variable as compared to the image of Cowley et al. (1993).

In order to compare the HRI intensities of RX J0527.8–6954 with those measured with the PSPC we determined the PSPC/HRI countrate ratio for this supersoft X-ray spectrum in the following two ways: 1) In an empirical approach we selected a few single white dwarfs (WDs) which have been observed with both, the PSPC and HRI, and derive a conversion factor PSPC/HRI = 7.8. This is thought to be an upper limit because such isolated WDs are less absorbed than SSS and thus sample even the lowest PSPC channels. 2) We have used the best fit model of the PSPC spectrum as input for computing the expected HRI countrate using its up-to-date response matrix and effective area. The result is sensitive to the temperature chosen and gives a ratio of PSPC/HRI = 7.7 (7.0) for a blackbody temperature of kT = 35 (40) eV. Thus, we adopt a ratio of 7.5.

### 2.4 The Lightcurve

Fig. 2 shows the X-ray lightcurve of RX J0527.8–6954 (i.e. the background subtracted counts divided by the effective exposure time) as deduced from 19 *ROSAT* PSPC pointings, six HRI pointings and the All-Sky-Survey data between 1990 and 1994. Two main features can be recognized immediately from this overall 5 year lightcurve: 1) The source has exponentially declined in X-ray intensity since its first *ROSAT* observations in 1990 ($\tau=1.7\pm0.1$ yrs), and 2) there is considerable scatter in the decline which is larger than our estimate of the remaining systematic errors in correcting for the effects of the window support structure.

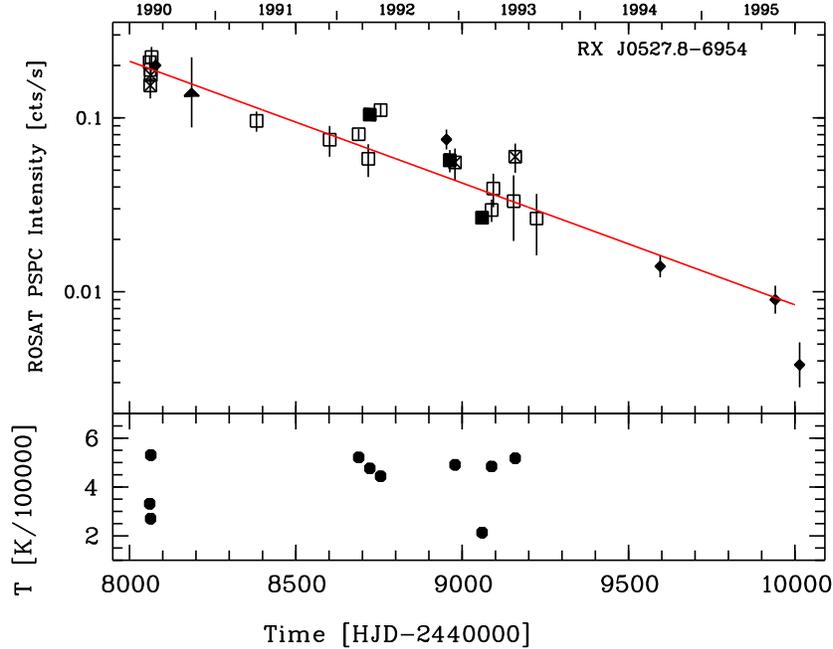

**Fig. 2.** X-ray lightcurve of RX J0527.8–6954 over the past 5 years as observed with the *ROSAT* satellite. The triangle marks the mean of the scanning observation during the All-Sky-Survey, squares mark PSPC observations (open symbols for off-axis angles larger than 15′ and affected by the window support structure, crossed squares for off-axis angles larger than 15′ and no obscuration by ribs, and filled symbols for off-axis angles smaller than 15′), and filled hexagons denote HRI observations with the countrates transformed to PSPC rates (see text). Systematic errors (not included in the plotted error bars) are largest for observations marked with open squares and might reach a factor of 2. The solid line is an exponential with $\tau=1.7$ years. The lower panel shows the best-fit blackbody temperature (while $N_H$ was fixed) derived from the PSPC observations. The low temperature point at HJD = 9060 (ROR 400298a) and the spread of best-fit temperatures during the All-Sky-Survey are possibly caused by inadequately corrected gain differences at various off-axis angles.

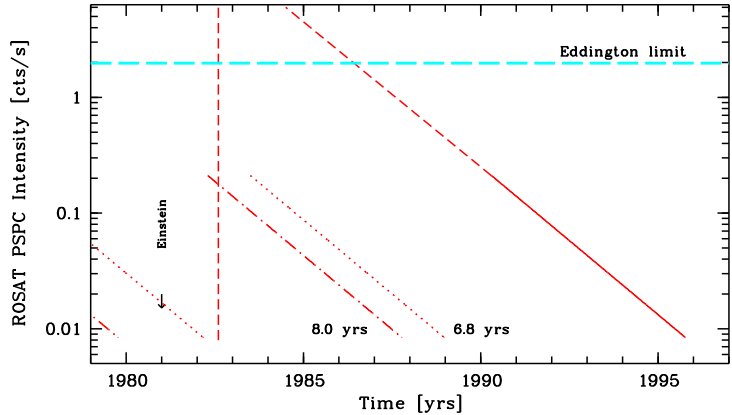

**Fig. 3.** Schematic presentation of the possible timescales using the $\tau=1.7$ yrs exponential of Fig. 2 and the *Einstein* upper limit. For a repeating source with sawtooth profile the repitition time scale is constrained to 6.8–8 yrs.

Surprisingly, the last HRI observation in October 1995 shows RX J0527.8–6954 to be considerably fainter than the extrapolation of the exponential decline. The X-ray intensity has dropped by a factor of 2.5 within two months. We have checked the housekeeping data of this pointing for any anomalous behaviour in detector properties or background variation - with negative result. Unfortunately, the only steady X-ray source (the SNR N132D) is at 20′ off-axis and thus cannot be used as an observational calibration. With the data products of this observation being so recent, further checks of the instrument performance are certainly necessary. If this intensity drop indeed is real, one might speculate on having caught the source during the switch-off.

The total X-ray intensity amplitude between maximum and minimum observed *ROSAT* intensity is a factor of 50 within these 5 years. This is about a factor five larger than the estimated amplitude deduced from the *Einstein* non-detection. At the present X-ray intensity (and during all the last two years) the source would have been invisible again for the *Einstein* observatory.

The decay time up to now is >5.5 yrs. Extrapolating this decay law back and using the *Einstein* upper limit allows to constrain the recurrence time to 6.8–8.0 yrs, if the rise time is very fast (see Fig. 3). The other alternative, a recurrence time of 13.5–15 yrs, would imply that the source exceeded the Eddington limit by a factor of a few for several years.

Though the number of counts detected during the individual PSPC pointings is mostly rather low, we investigated the possibility of X-ray spectral changes during the decline. First, we kept the absorbing column fixed at its galactic value and determined the temperature being the only fit parameter. We find no systematic trend of a temperature decrease (lower panel of Fig. 2). Second, we kept the temperature fixed (at 40 eV in the first run and at the best fit value of the two parameter fit in the second run) and checked for changes in $N_H$, again finding no correlation.

## 3  Discussion

The most popular model of SSS involves steady nuclear burning on the surface of an accreting WD (van den Heuvel et al. 1992). If $\dot M$ exceeds a certain value that depends on the WD mass and other physical parameters, hydrogen burning on the WD surface is stable.

There are several possible phenomena which might explain completely or partially the X-ray variability of RX J0527.8–6954:

1. Changes in $\dot M$ within the small range which is necessary for stable hydrogen burning. The luminosity amplitude in this case can be only a factor 2.3–2.7 (see Fig. 9 of Fujimoto 1982 and Figs. 2 and 8 of Iben 1982). However, since the nuclear luminosity depends mainly on the conditions of the burning envelope, the corresponding changes are expected on the accretion time scale, i.e. much longer than observed in RX J0527.8–6954. Therefore, such $\dot M$ changes are ruled out as the only cause of the exponential decline.
2. The atmospheric layers are expanding and thus the effective temperature decreases, while the bolometric luminosity remains constant (evolution along the horizontal track in the $\log(T)$- $\log(L)$ plane). The expansion could be caused by increased mass transfer from the secondary. This mechanism has already been suggested by Pakull et al. (1993) for RX J0513.9–6951. However, is is unlikely for RX J0527.8–6954 since (1) it works only near the Eddington rate and (2) no temperature decrease has been observed.
3. The gradual decline of the X-ray intensity also might suggest that we observe the decay phase after a shell flash. Thus, it is important to ask when the decay of RX J0527.8–6954 might have started. The recurrence time of hydrogen flashes is inversely proportional to the mass accretion rate and to the white dwarf mass. Since RX J0527.8–6954 has not been detected with *Einstein* observations, the shell flash must have occurred after 1981. Moreover, since with *ROSAT* we witness a decline from the beginning, the hydrogen-burning plateau phase (horizontal track in the H–R diagram) must have been short. Thus, even if the shell flash happened just after the *Einstein* observation then it would still need a considerable fine tuning of WD mass and accretion rate ($\dot M \leq 10^{-8} M_\odot$ yr$^{-1}$) for the WD to be on the declining portion of the evolutionary track within only 10 years. Moreover, a shell flash with such properties would be accompanied by an optical brightening of the object to $M_V \simeq -5$. At the LMC distance this would correspond to a visual brightness of 14th magnitude which very certainly would not have passed undetected.
4. The WD is cooling at constant radius after a weak flash. Due to the strong sensitivity of the *ROSAT* countrate on the temperature the X-ray amplitude of a factor of 10 does not necessarily translate into a factor of 10 change in bolometric luminosity (even if the latter is assumed to be dominated by the soft X-rays). Thus, the WD can remain within the stable burning range (with its factor 2–3 reduction in bolometric luminosity) while the concordant temperature decrease will shift the Wien tail out of the *ROSAT* window. If the cooling alone would account for the X-ray intensity decay, then the observed intensity amplitude corresponds to a temperature amplitude of at least two (depending on the absolute temperature). This translates into a cooling time of the order of five years. We can assume that the shortest time for the white dwarf cooling is only slightly longer than the time necessary to burn all the hydrogen envelope mass once there is no mass transfer at all, like it is usually supposed after a nova outburst (Starrfield, priv. comm.). This can be as short as 5 years for a white dwarf of 1 $M_\odot$ (Kato & Hachisu, 1994).

Thus, it seems likely that the variability of RX J0527.8–6954 is due an earlier hydrogen flash on a rather massive white dwarf (m$\geq$ 1.1$M_\odot$) accreting at a high rate ($10^{-7}$ $M_\odot$/yr) as described in Iben (1982) and Fujimoto (1982). In an empirical model of recurrent SSS, Kahabka (1995) related the envelope mass to the decay and recurrence times of these sources. The latter can be used in turn as observables to determine the WD mass and its accretion rate. From a preliminary analysis of the X-ray decline of RX J0527.8–6954 Kahabka (1995) adopted a value of 5 yrs for the time for return to minimum after a flash, and a recurrence time of 10 yrs. These numbers have changed only moderately with the herewithin reported results. With the new timescales we derive $M_{WD}$=1.2–1.35 $M_\odot$ and $\dot{M}_{accr}$=2–7$\times 10^{-7}$ $M_\odot$/yr. The rather high WD mass is also compatible with the temperature of 5–6$\times 10^5$ K and the derived luminosity.

*Acknowledgements:* JG is supported by the Deutsche Agentur für Raumfahrtangelegenheiten (DARA) GmbH under contract FKZ 50 OR 9201. We are extremely grateful to Y.-H. Chu for providing the data of the October 1995 HRI observation. We thank C. Motch for valuable comments. The *ROSAT* project is supported by the German Bundesministerium für Bildung, Forschung, Wissenschaft und Technologie (BMBW/DARA) and the Max-Planck-Society.